\documentclass[conference]{IEEEtran}
\IEEEoverridecommandlockouts
\usepackage{amsmath,amssymb,amsfonts}
\usepackage[T1]{fontenc}
\usepackage{graphicx}
\usepackage{textcomp}
\usepackage{booktabs}
\usepackage{multirow}
\usepackage{enumitem}
\usepackage{subcaption}

\usepackage{longtable}  

\usepackage{xcolor}
\usepackage{tikz}

\usepackage[
 backend=biber
,style=ieee
,minbibnames=1
,maxbibnames=2
,maxcitenames=1
,mincitenames=1
,eprint=true
,doi=true
,isbn=false
,url=true
]{biblatex}
\bibliography{mybib} 
\AtBeginBibliography{\footnotesize}

\DeclareSourcemap{
  \maps{
    \map{
      \step[fieldset=month, null]
      \step[fieldset=address, null]
      \step[fieldset=location, null]
      \step[fieldset=publisher, null]
      \step[fieldset=url, null]
      \step[fieldset=isbn, null]
    }
  }
}

\def\BibTeX{{\rm B\kern-.05em{\sc i\kern-.025em b}\kern-.08em
    T\kern-.1667em\lower.7ex\hbox{E}\kern-.125emX}}

\usepackage{algorithm}
\usepackage[noend]{algpseudocode}
\algnewcommand{\LineComment}[1]{\State \(\triangleright\) #1}
\algnewcommand{\algorithmicand}{\textbf{ and }}
\algnewcommand{\algorithmicor}{\textbf{ or }}
\algnewcommand{\OR}{\algorithmicor}
\algnewcommand{\AND}{\algorithmicand}
\algnewcommand{\var}{\texttt}

\makeatletter

\usepackage{cleveref}
\Crefname{figure}{Fig.}{Figs.}

\usepackage{flushend}

\begin{document}

\title{Early Detection of Network Service Degradation: An Intra-Flow Approach}



\author{
    \IEEEauthorblockN{Balint Bicski\IEEEauthorrefmark{1}\IEEEauthorrefmark{3} and Adrian Pekar\IEEEauthorrefmark{1}\IEEEauthorrefmark{2}\IEEEauthorrefmark{3}}
    \IEEEauthorblockA{
        \IEEEauthorrefmark{1}Department of Networked Systems and Services, Faculty of Electrical Engineering and Informatics,\\ Budapest University of Technology and Economics, M\H{u}egyetem rkp. 3., H-1111 Budapest, Hungary.\\
        \IEEEauthorrefmark{2}HUN-REN-BME Information Systems Research Group, Magyar Tud\'{o}sok krt. 2, 1117 Budapest, Hungary.\\
        \IEEEauthorrefmark{3}CUJO LLC, Budapest, Hungary.\\
        Email: balint.bicski@edu.bme.hu, apekar@hit.bme.hu
    }
}

\maketitle

\begin{abstract}
This research presents a novel method for predicting service degradation (SD) in computer networks by leveraging early flow features. Our approach focuses on the observable (O) segments of network flows, particularly analyzing Packet Inter-Arrival Time (PIAT) values and other derived metrics, to infer the behavior of non-observable (NO) segments. Through a preliminary evaluation, we identify an optimal O/NO split threshold of 10 observed delay samples, balancing prediction accuracy and resource utilization. 
Evaluating models including Logistic Regression, XGBoost, and Multi-Layer Perceptron, we find XGBoost outperforms others, achieving an F\textsubscript{1}-score of 0.74, balanced accuracy of 0.84, and AUROC of 0.97. Our findings highlight the effectiveness of incorporating early flow features and the potential of our method to offer a practical solution for monitoring network traffic in resource-constrained environments. 
\end{abstract}


\begin{tikzpicture}[remember picture,overlay]
\node[anchor=north, align=center, text=red, font=\small, yshift=-.6cm] at (current page.north) {This version of the paper was accepted for presentation at the 2024 International Conference on Network and Service Management (CNSM 2024).};
\end{tikzpicture}

\section{Introduction}

In the era of relentless digitization, residential networks are fundamental to digital activities, from streaming videos to real-time gaming. As the demand for uninterrupted, high-quality digital experiences grows, so does the need for resilient networks that can consistently meet these expectations~\cite{Cito2014,Vuleti2020}. However, service degradation (SD), characterized by reduced network performance, poses a significant challenge. Factors like network congestion, inefficient data routing, or external interferences can lead to increased latency, buffering, or even outages~\cite{Casas2016}. Therefore, effective SD management is crucial for seamless digital experiences.

Research on SD detection has evolved over nearly two decades, exploring various metrics and methods. Early work analyzed round-trip time deviations to predict Internet SDs~\cite{1217281}. Subsequent studies have utilized end-to-end delay and loss measurements to assess network quality~\cite{ABDELKEFI201430} and developed real-time packet loss monitoring systems~\cite{9873985}. Machine learning approaches have also been explored; for example, deep neural networks have been used to predict throughput and flow duration in campus networks~\cite{9201025}, highlighting their potential in SD prediction.

Latency has been identified as a crucial SD indicator, particularly in applications like video streaming and cloud gaming~\cite{10211223, 10041952}. In cloud computing environments, methods to detect inter-VM interference and estimate performance degradation using multi-variable regression models have been proposed~\cite{PONS202313}. Additionally, controlled SD strategies have been introduced to manage network overloads, balancing service quality and resource utilization~\cite{7875717, 8354697}.

Our research builds on these insights by focusing on SD detection in residential LAN environments, where network devices often have limited computational capabilities~\cite{Zulfiqar2023}. These devices typically use a dual-path architecture: a fast path for rapid hardware-based forwarding and a slow path for CPU-based processing~\cite{Molero2018,Cerovic2018,Zulfiqar2023}. Once specific criteria, like the completion of a TCP handshake, are met, flows transition from the CPU (slow path) to hardware (fast path), reducing the CPU workload but limiting detailed, packet-level monitoring~\cite{Cerovic2018}. Consequently, flows are split into observable (O) and non-observable (NO) segments, presenting a challenge in detecting SD events in the NO segments after offloading to hardware accelerators.

To address this challenge, we propose leveraging early flow features from the O segments to predict the status of the NO segments, thereby identifying potential SD in residential networks. Our approach, termed \textit{intra-flow service degradation detection}, uses early flow characteristics, such as Packet Inter-Arrival Time (PIAT), to gain insights into network performance. We evaluate various models, including Logistic Regression, XGBoost, and Multi-Layer Perceptron (MLP), to determine their effectiveness in predicting SD events from the O segments. Results show that XGBoost, with an O/NO split threshold of 10 observed delays, offers the best performance---achieving an F\textsubscript{1}-score of 0.74, balanced accuracy of 0.84, and AUROC of 0.97---balancing prediction accuracy with resource utilization.

Our main contributions in this paper are as follows:
\begin{itemize}
    \item A novel approach for predicting SD in network flows using early flow features to infer behavior in NO parts.
    \item Introduction and evaluation of intra-flow SD detection, leveraging information within a flow for prediction in resource-constrained environments.
    \item Identification of an optimal O/NO split threshold, balancing prediction accuracy and practical implementation in network devices with hardware offloading.
\end{itemize}

The rest of this paper is organized as follows: 
\Cref{sec:methodology} presents our method to identify SD events. \Cref{sec:evaluation} evaluates the predictive power of O parts of network flows in a binary classification setting. \Cref{sec:conclusion} concludes the paper.

\section{Methodology}
\label{sec:methodology}









\subsection{Vertical Separation of Delays}
\label{sec:vertical_separation}

\begin{figure}[!t]
\centering
\includegraphics[width=.9\linewidth]{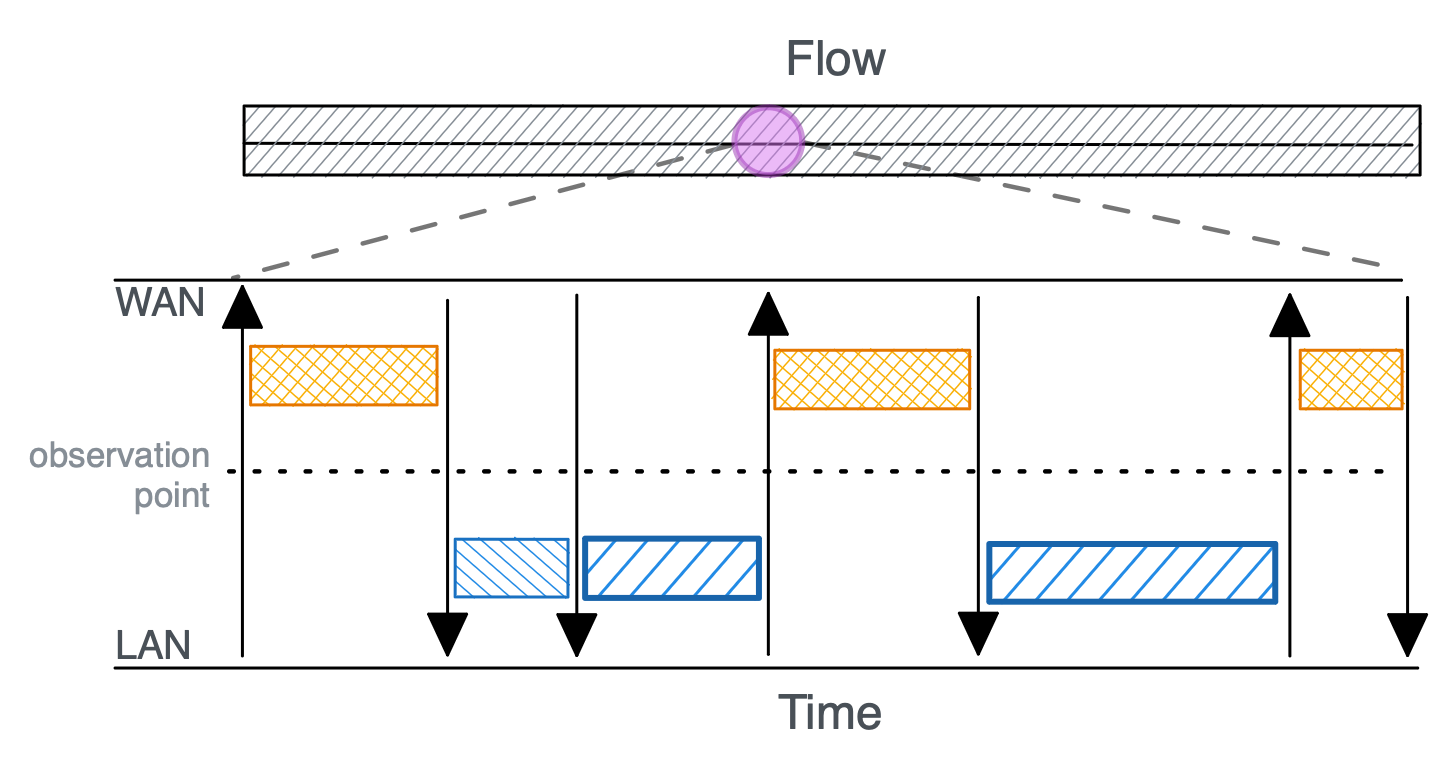}
\caption{Visual representation of network traffic flow showing the vertical separation between WAN and LAN delays.}
\label{fig:vert_sep} 
\end{figure}

Network flow packets can be categorized into LAN and WAN directions, with PIATs measured for each. 
PIAT quantifies the time interval between the arrivals of two consecutive packets within a network flow at the flow meter. We derive LAN-side delays (and jitter) from this metric that serve as foundational metrics of our methodology~\cite{3456588}.
The concept of \textit{vertical separation} focuses on isolating LAN-side delays to accurately assess local network conditions, avoiding variability from WAN influences. \Cref{fig:vert_sep} illustrates this separation, representing packet directionality as arrows and showing LAN PIATs in blue and WAN delays in orange.

Assuming that delays caused by local endpoints are negligible compared
to those induced by broader network conditions and that the response to the last packet in a burst is instantaneous
(a condition typically met with TCP traffic) the LAN PIATs marked with a thick blue border in \Cref{fig:vert_sep} correspond to the latency introduced by traversing the LAN, thus defining the LAN-side delays that we use in our analysis.


\subsection{Horizontal Separation of Flows}

Flows in a network can be split into \textit{Observable} (O) and \textit{Non-observable} (NO) segments based on a threshold \( \theta \), which defines the number of packets monitored by the software layer before offloading to hardware. If a flow’s packet count \( p \) is:
\[
\text{State of Flow} = 
\begin{cases} 
\text{Observable} & \text{if } p \leq \theta \\
\text{Non-observable} & \text{if } p > \theta 
\end{cases}
\]

\begin{figure}[!t]
\centering
\includegraphics[width=.9\linewidth]{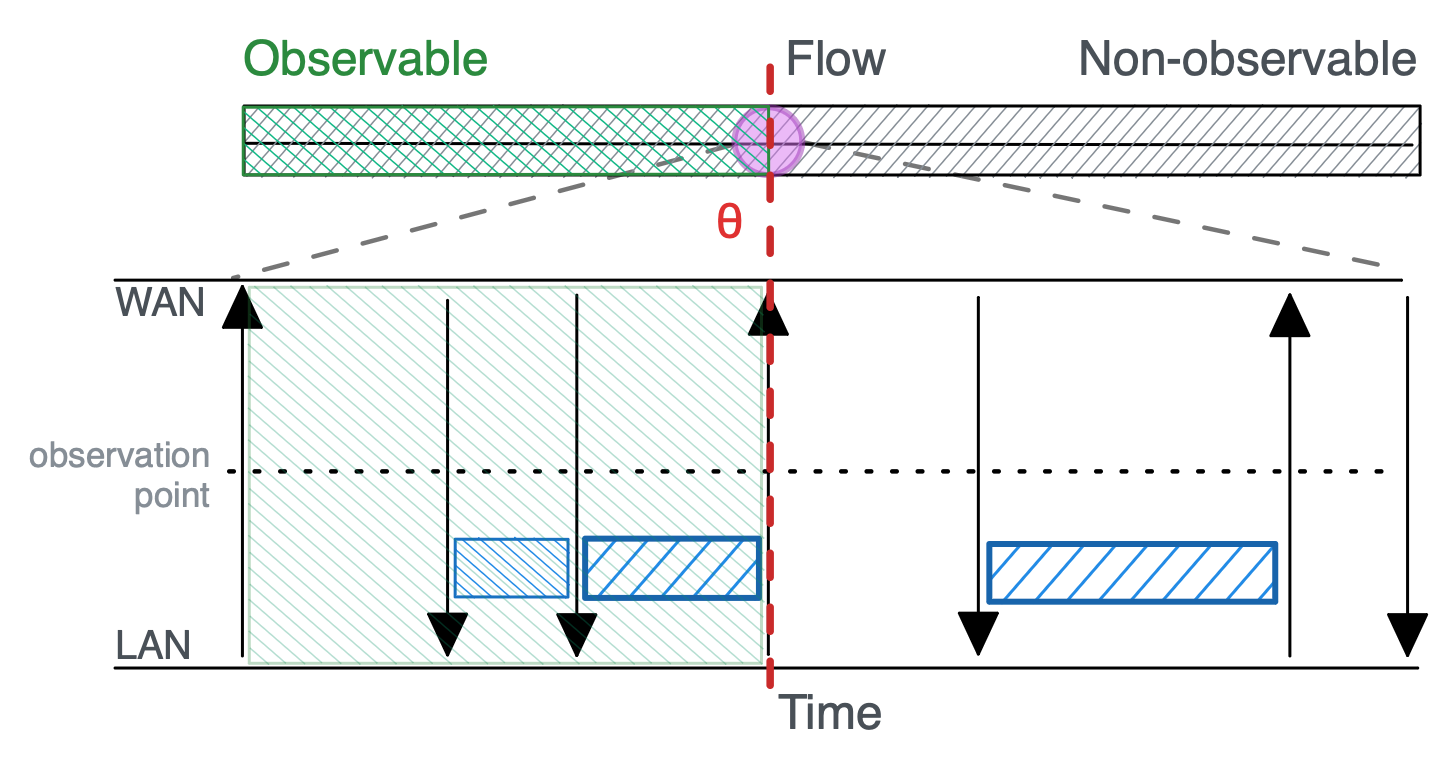}
\caption{Illustration of the horizontal separation of a network flow into O and NO segments.}
\label{fig:hor_sep} 
\end{figure}

For \( ||f_i|| \leq \theta \), all packets are observed; for \( ||f_i|| > \theta \), the flow is divided into \( O_p = \{p_1,\dots,p_\theta\} \) (observed) and \( NO_p = \{p_{\theta+1},\dots,p_{||f_i||}\} \) (unobserved). This \textit{horizontal separation}, visualized in \Cref{fig:hor_sep}, highlights the point where a flow transitions from software to hardware processing, marked by the threshold \( \theta \).

\subsection{Refining Observability of LAN Delays}

The \( \theta \) parameter does not ensure a consistent number of observable LAN delays due to the exclusion of WAN delays and the occurrence of packet bursts. To refine the focus on LAN delays, we introduce a threshold \( m \) specific to LAN delays. The O part of the flow consists of LAN delays up to \( m \), while delays beyond this are considered non-observable. If a flow has fewer delays than \( m \), it is fully observable.

\subsection{Heuristic Detection of Service Degradation Events}

\begin{figure}[!t]
    \centering
    \includegraphics[width=.9\linewidth]{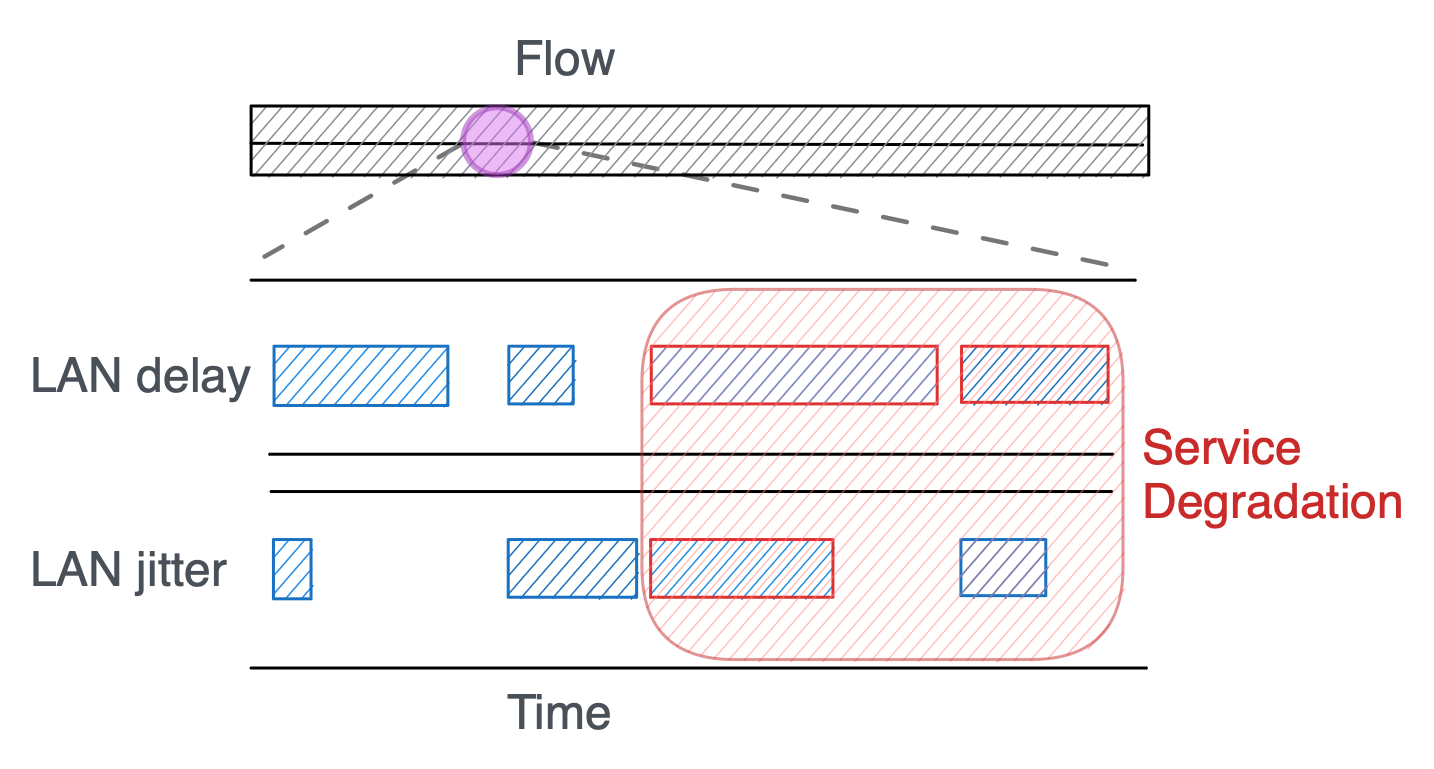}
    \caption{An SD event in a flow with a MSL of 2.}
    \label{fig:SD_event} 
\end{figure}

With refined observability, SD events can be detected using a heuristic based on latency and jitter behavior over time~\cite{3456588}. Different application types have unique requirements for extreme delay and jitter occurrences, defined by a \textit{Minimum Sequence Length} (MSL) of such events. An SD event begins with an extreme delay and jitter followed by a sequence of extreme delays. \Cref{fig:SD_event} shows an SD event with MSL of 2.

\subsection{Handling Split SD Events}

SD events can span the O and NO parts of flows, requiring careful handling to maintain detection accuracy. Three scenarios arise during horizontal separation:
\begin{itemize}
    \item The SD event is entirely within the O part (\( SD_{\text{end}} < O_{\text{end}} \)).
    \item The SD event is entirely within the NO part (\( SD_{\text{start}} > O_{\text{end}} \)).
    \item The SD event splits between O and NO parts (\( SD_{\text{start}} < O_{\text{end}} \) and \( SD_{\text{end}} > O_{\text{end}} \)).
\end{itemize}

\begin{figure}[!t]
    \centering
    \includegraphics[width=.56\columnwidth]{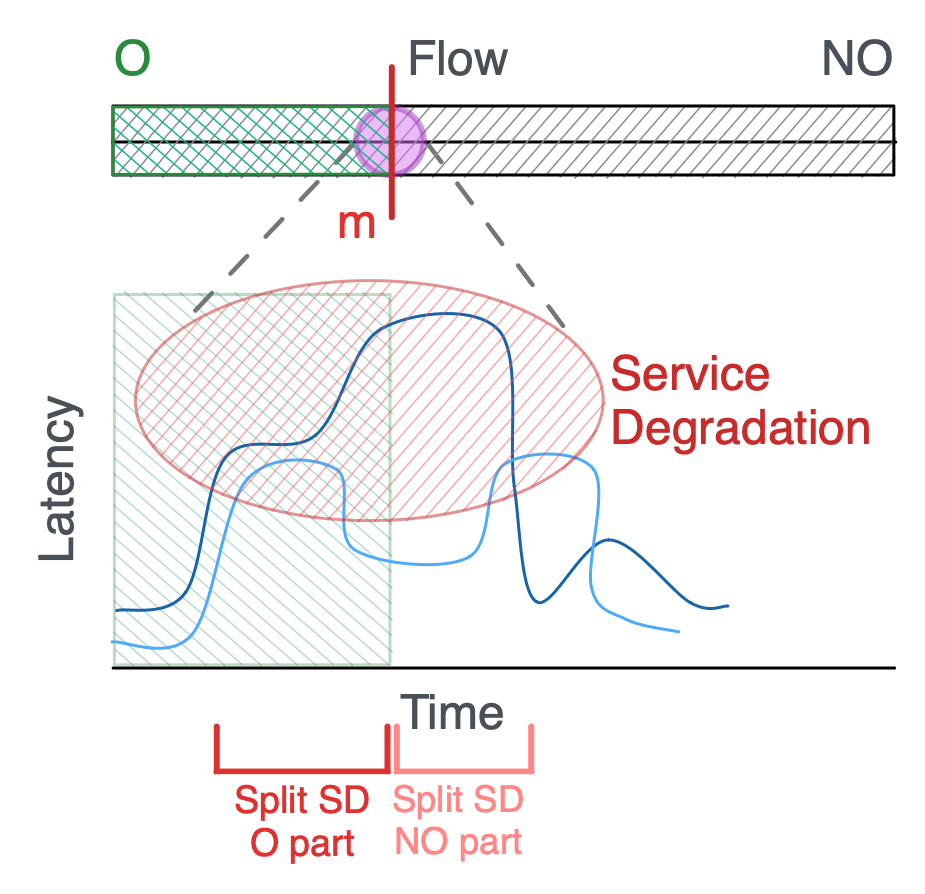}
    \caption{Illustration of a split SD event in a flow.}
    \label{fig:split_sd}
\end{figure}

In the last case, in order to preserve potentially
crucial information we chose to keep  these
events by splitting them in two. \Cref{fig:split_sd} illustrates this scenario.

\begin{figure}[!ht]
    \centering
    \includegraphics[width=.65\columnwidth]{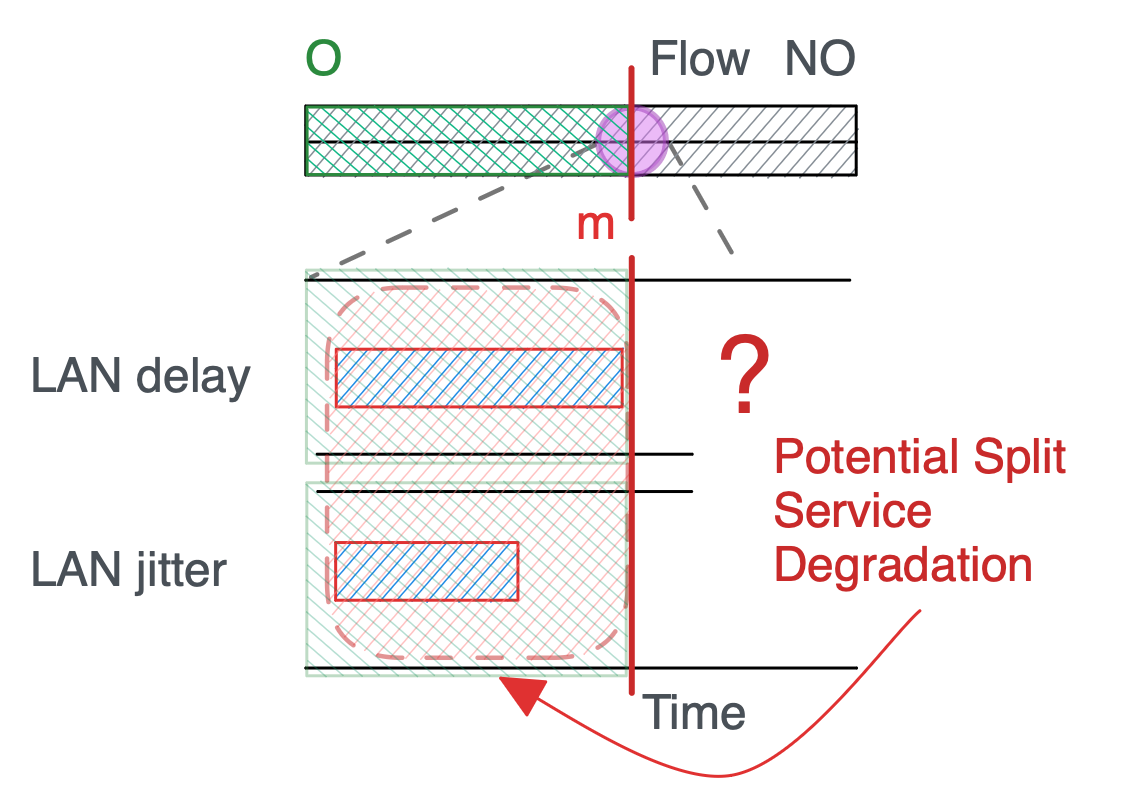}
    \caption{Illustration of a potential split SD event.}
    \label{fig:potential_split_sd}
\end{figure}

However, this process makes these split events indistinguishable from apparent SD events that are shorter than MSL (i.e. they do not develop into real SD events) and either cross the O/NO boundary or end right at the end of the O part. Such a split is depicted in \Cref{fig:potential_split_sd}. Aiming for consistency, we marked these the same way as real split SD events.

To analyze the splits, we calculate the \textit{split SD ratio} as:
\[
f_{\text{split SD ratio}} = \frac{f_{|\text{partial SD event}|}}{f_{\text{MSL}}},
\]
where \( f \) is a flow. This ratio helps quantify how close a partial event is to meeting the SD event criteria.

\subsection{Intra-Flow Service Degradation Detection}

\begin{figure}[!t]
    \centering
    \includegraphics[width=\columnwidth]{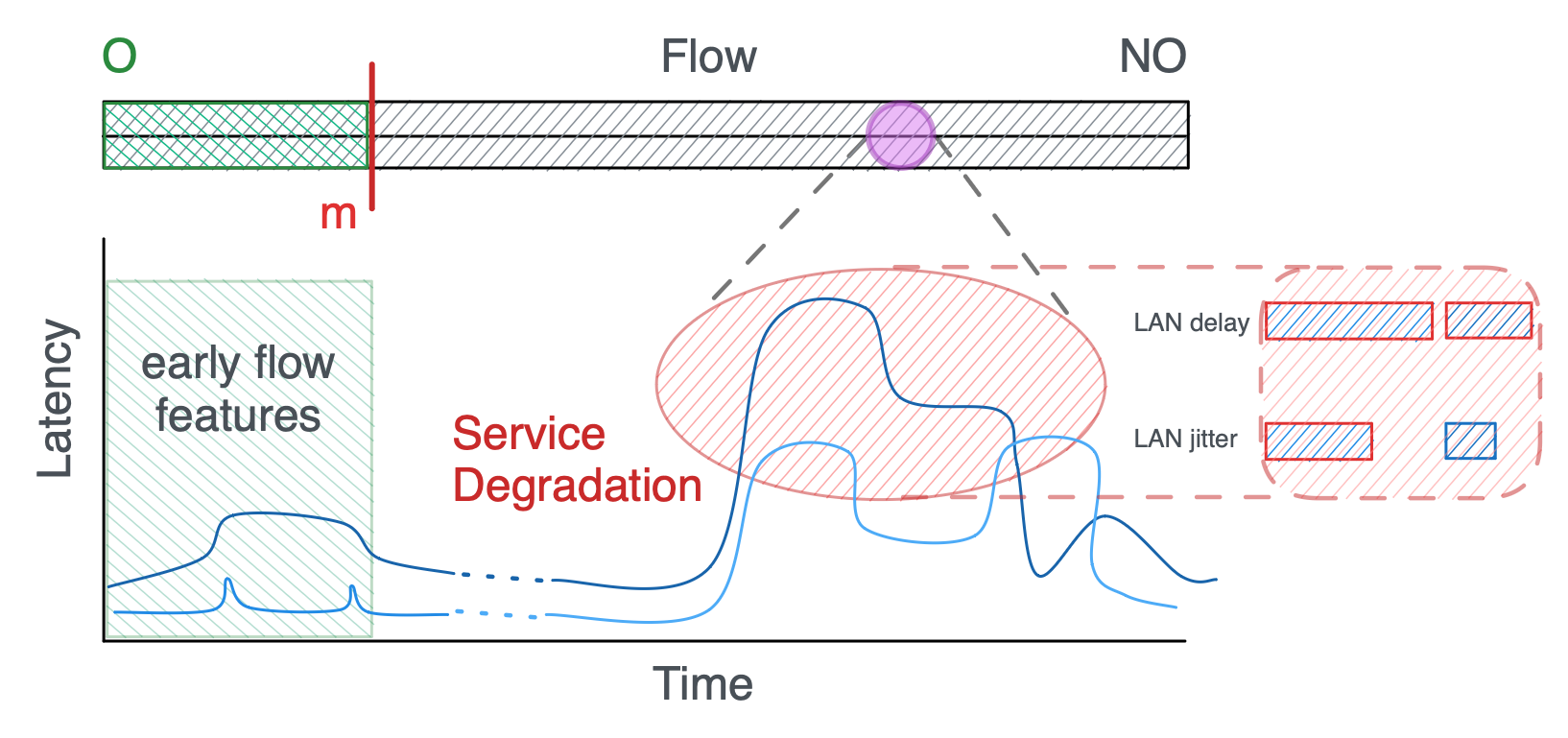} 
    \caption{Intra-flow SD detection.}
    \label{fig:intra_flow_analysis}
\end{figure}

Accurate SD detection in resource-constrained edge devices requires leveraging the observable (O) flow segments to infer the state of the non-observable (NO) segments. We propose an \textit{intra-flow service degradation detection} method that uses early flow features from the O part to predict SD events in the NO part. \Cref{fig:intra_flow_analysis} illustrates this approach.

The goal is to perform a binary classification using the O segment data to predict SD presence in the NO segment, offering a solution tailored to the limitations of edge devices that cannot continuously monitor entire flows.

\section{Preliminary Evaluation and Performance Analysis}
\label{sec:evaluation}

We assess the potential of the O parts of flows to predict SD events in the NO parts using binary classification models to determine SD event presence or absence in the NO part.

\subsection{Dataset}
We use the dataset from \cite{3456588,sdd-github}, which consists of TCP network flows from five consecutive days in a university dormitory network. The dataset includes:
\begin{itemize}
    \item Aggregated flow features and packet-level characteristics for the first 255 packets of each flow.
    \item LAN delay and jitter values extracted using vertical separation (\Cref{sec:vertical_separation}).
    \item Marked SD events with application-specific MSL thresholds.
\end{itemize}

The analysis focuses on flows measured at \textit{Location 2} for result transferability.

\subsection{Data Preparation}

Data from Monday to Wednesday was used for training, and Thursday to Friday for testing. Only flows with an O/NO split were included. Input features include:
\begin{itemize}
    \item Statistical measures (min, max, median, mean, std) of delays and jitters in the O part.
    \item Individual delay and jitter values, SD event counts, and attributes of the longest SD event in the O part.
    \item Application, category, location, connection type, and $f_{\text{split SD ratio}}$.
\end{itemize}

We performed one-hot encoding on categorical variables and used Standard Scaler for scaling. \Cref{tab:training_testing_data} summarizes the input sizes and feature counts for different O/NO splits.

\begin{table*}[ht]
\scriptsize
\centering
\caption{Train and Test Sizes with Input Feature Counts for Different O/NO Splits}
\label{tab:training_testing_data}
\begin{tabular}{c|c|c|p{0.6\textwidth}}
\toprule
\textbf{O/NO Split} & \textbf{Train Size} & \textbf{Test Size} & \textbf{Input Feature Count} \\ \midrule
\textbf{5}  & 905,407 flows & 507,359 flows & 141 features (including 6 app. cat. names, 101 app. name, 9 location, 2 connection type OH encoded features) \\ 
\textbf{10} & 264,183 flows & 154,021 flows & 140 features (including 6 app. cat. names, 90 app. name, 9 location, 2 connection type OH encoded features) \\ 
\textbf{15} & 168,463 flows & 99,290 flows  & 144 features (including 6 app. cat. names, 84 app. name, 9 location, 2 connection type OH encoded features) \\ 
\textbf{20} & 122,814 flows & 73,043 flows  & 152 features (including 6 app. cat. names, 82 app. name, 9 location, 2 connection type OH encoded features) \\ \bottomrule
\end{tabular}
\end{table*}

\begin{figure*}[ht]
    \centering
    \begin{subfigure}[b]{0.49\linewidth}
        \includegraphics[width=\linewidth, trim={0 .8cm 0 .2cm}, clip]{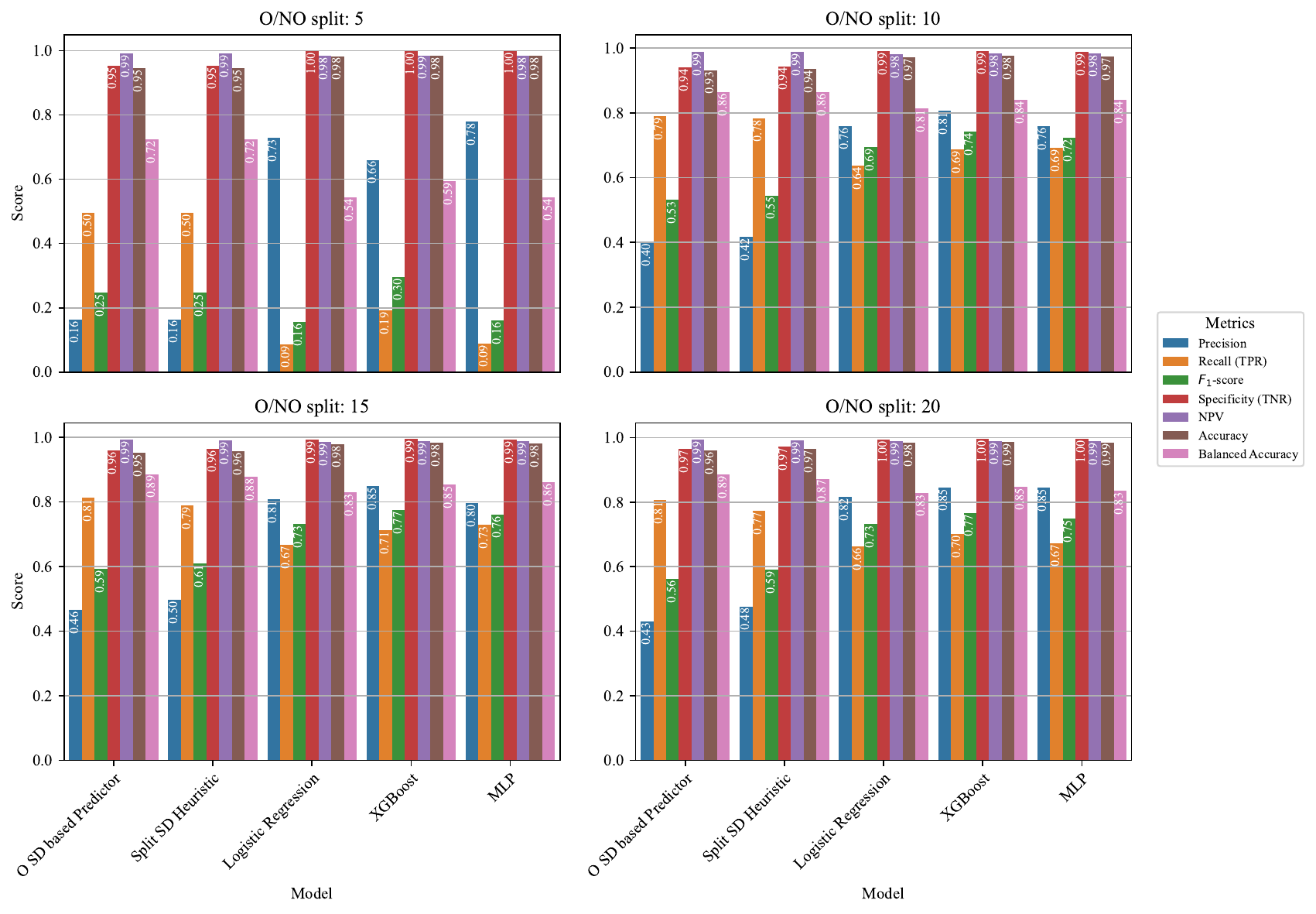}
        \caption{Classification metrics for NO SD presence prediction.}
        \label{fig:class_metrics}
    \end{subfigure}
    \hfill
    \begin{subfigure}[b]{0.49\linewidth}
        \includegraphics[width=\linewidth, trim={0 .4cm 0 1.1cm}, clip]{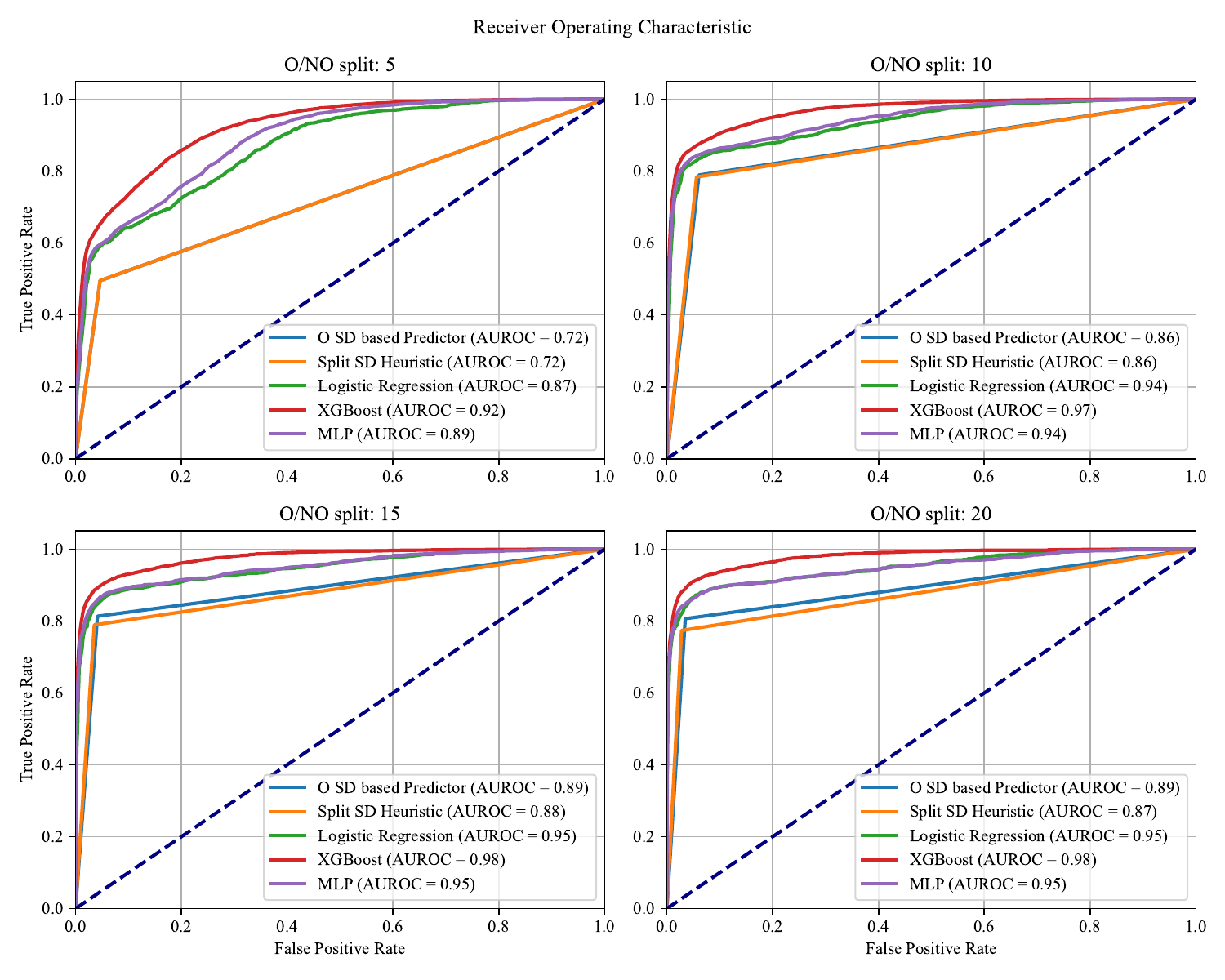}
        \caption{ROC curves with AUROC metrics for NO SD presence prediction.}
        \label{fig:class_roc}
    \end{subfigure}
    \caption{Performance at different O/NO split thresholds.}
    \label{fig:class_metrics_combined}
\end{figure*}

\subsection{Experimental Design and Evaluation Metrics}
We evaluated the following models:
\begin{itemize}
    \item Baseline predictors (null, all-true, random),
    \item Heuristic models (SD-Based Predictor, Split SD Metric-Based Predictor),
    \item Machine learning models (Logistic Regression, XGBoost, Multi-Layer Perceptron (MLP)).
\end{itemize}

Grid Search and 5-fold cross-validation were used to optimize model parameters. Performance was evaluated using metrics: \textit{precision}, \textit{recall}, \textit{F\textsubscript{1}-score}, \textit{specificity}, \textit{NPV}, \textit{accuracy}, and \textit{balanced accuracy}.


\subsection{Classification Performance}
\label{sec:classification}

\Cref{fig:class_metrics_combined} shows the achieved results for the selected O/NO threshold splits. \Cref{fig:class_metrics} compares the classification metrics, while \Cref{fig:class_roc} depicts the Receiver Operating Characteristic (ROC) curves and corresponding areas under the curves (AUROC).

\subsubsection{Overall Performance}


All trained models significantly outperformed the random, all-true, and all-null predictors. The null predictor achieved over 95\% accuracy but only 0.5 balanced accuracy, reflecting a high prevalence of flows without SD events in the NO part.


All models at every threshold showed excellent specificity, NPV, and accuracy (over 95\%, many approaching 100\%), indicating their effectiveness in correctly classifying flows without SD in the NO part. This suggests consistent behavior for flows with no SD events in the O part, nor split SDs, as evidenced by high TNR and NPV values. The high accuracy reflects the data imbalance toward these types of flows. However, models struggled to correctly identify flows with non-observable SDs, where these events were less frequent.

\subsubsection{Model Comparison}


Heuristic models (SD Based Predictor and Split SD Heuristic) showed moderate effectiveness with similar prediction results across all thresholds. They achieved the highest recall and balanced accuracy (50\% and 72\% at threshold 5, and around 0.8 and over 0.85 at higher thresholds). However, they underperformed in precision and F\textsubscript{1}-score compared to other models, indicating that while many flows with non-observable SD events have SD in their O parts, many flows with SD in the O part or guessed split SD events do not develop SD in the NO part.


The sophisticated models (Logistic Regression, XGBoost, and MLP) had much higher precision but lower recall, especially at threshold 5. For instance, XGBoost reached 66\% precision (over 72\% for the others) but had a recall of only 19\%, and just 9\% for Logistic Regression and MLP. This resulted in low F\textsubscript{1}-scores (30\% for XGBoost and about half for the others) and balanced accuracy around 55-60\%, suggesting that lower thresholds do not provide enough information to predict the more volatile behavior in the NO part effectively.

\subsubsection{Impact of O/NO Split Threshold}


At the O/NO split threshold of 10, prediction metrics were more balanced for all models. The heuristics maintained high recall (78\%) and balanced accuracy (86\%) but had limited improvement in precision (under 45\%). The sophisticated models achieved more balanced results, with all metrics above 64\%. XGBoost and MLP had a recall of 69\% and balanced accuracy of 84\%, with XGBoost achieving over 80\% precision and the highest F\textsubscript{1}-score (74\%).


Beyond threshold 10, splits of 15 and 20 showed only marginal gains. XGBoost reached 85\% precision and a 77\% F\textsubscript{1}-score, but further increases led to minor or even decreased improvements, likely due to fewer flows being considered. This indicates that higher thresholds offer no significant advantages.

\subsubsection{ROC Curve Analysis}

The ROC curves show similar behavior to the patterns identified earlier. The simple heuristics demonstrate clear inferiority to the other models, even at the O/NO split threshold of 5 (with AUROC scores of 72\% compared to over 87\% for the other models). The superior robustness of the XGBoost model is evident at every threshold, exceeding 0.9 in AUROC at the O/NO split of 5, reaching 0.97 by threshold 10 and 0.98 by 15. 

Logistic Regression and MLP exhibit similar patterns with highly correlated ROC curves and nearly identical AUROC values. The lack of significant improvement beyond threshold 10 is also evident in this figure.

\subsubsection{Summary of Findings}


All models outperformed the baseline predictors, underscoring the effectiveness of our approach. The XGBoost model offered the best balance between high precision, recall, and balanced accuracy, making it the most suitable for predicting non-observable SD events. Threshold 10 emerged as the optimal O/NO split point, offering a strong trade-off between predictive performance and computational efficiency. Lower thresholds, like 5, lacked sufficient information for reliable predictions, while higher thresholds, such as 15 and 20, provided only marginal improvements, leading to diminishing returns.

\subsubsection{Limitations}

Our study, while promising, has certain limitations. The analysis was conducted on a specific network environment, and further validation in diverse settings would enhance the generalizability of our findings. The accuracy of predictions may vary depending on the nature of the flow and the specific characteristics of SD events. While our model performs well in predicting the presence of SD events, predicting their exact characteristics (such as duration or severity) remains a challenge. Additionally, the method's effectiveness is influenced by the choice of the O/NO split threshold and the features selected for analysis. Despite these limitations, our intra-flow SD detection approach offers a novel and resource-efficient method for maintaining network quality in challenging environments.

\section{Conclusion}
\label{sec:conclusion}

This paper introduced a novel method for predicting service degradation (SD) in networks by leveraging early flow features, particularly packet inter-arrival times and related metrics, from the observable (O) segments to infer the behavior in non-observable (NO) segments. We identified an optimal O/NO split threshold of 10 observed delays, which provides a balance between prediction accuracy and resource efficiency. 

Our results showed that the XGBoost model performed best, achieving an F\textsubscript{1}-score of 0.74, a balanced accuracy of 0.84, and an AUROC of 0.97. These findings underscore the potential of using early flow features for effective SD detection in resource-constrained environments. 
Future work will explore the extension of this approach to detect other network anomalies and validate the model across different network environments and traffic types to enhance its robustness and applicability.

\section*{Acknowledgment}

Supported by the János Bolyai Research Scholarship of the Hungarian Academy of Sciences. Supported by the ÚNKP-23-5-BME-461 New National Excellence Program of the Ministry for Culture and Innovation from the source of the National Research, Development and Innovation Fund. 

\printbibliography

\end{document}